\documentclass[preprintnumbers,amsmath,amssymb, prd, nofootinbib, eqsecnum]{revtex4}
\usepackage{graphicx}% Include figure files
\usepackage{amsmath,amssymb,graphics,epsfig}
\usepackage{dcolumn}% Align table columns on decimal point
\usepackage{epsf}
\usepackage{epstopdf}
\usepackage {amssymb}
\newcommand{\nc}{\newcommand}
\nc{\ba}{\begin{eqnarray}}
\nc{\ea}{\end{eqnarray}}

\def\be{\begin{equation}}
\def\ee{\end{equation}}
\def\ba{\begin{eqnarray}}
\def\ea{\end{eqnarray}}

\begin{document}

\title{ $ \delta N$  Formalism in Anisotropic Inflation and \\
Large Anisotropic Bispectrum and Trispectrum }

\author{Ali Akbar Abolhasani$^{1}$}
\email{abolhasani-AT-ipm.ir}
\author{Razieh Emami$^{1}$}
\email{emami-AT-ipm.ir}
\author{Javad T. Firouzjaee$^{1, 2}$}
\email{j.taghizadeh.f-AT-ipm.ir}
\author{Hassan Firouzjahi$^{2}$}
\email{firouz-AT-mail.ipm.ir}

\affiliation{$^1$School of Physics, Institute for Research in
Fundamental Sciences (IPM),
P.~O.~Box 19395-5531,
Tehran, Iran}

\affiliation{$^3$School of Astronomy, Institute for Research in
Fundamental Sciences (IPM),
P.~O.~Box 19395-5531,
Tehran, Iran}

\begin{abstract}
\vspace{0.3cm}
We present a consistent $\delta N$ formalism for curvature perturbations in anisotropic cosmological backgrounds. We employ our $\delta N$ formalism to calculate the power spectrum, the bispectrum and the trispectrum in models of anisotropic inflation with the background gauge fields in Bianchi I universe.  Our results coincide exactly with the recent results obtained from in-in formalism. To satisfy the observational constraints the anisotropies generated on power spectrum are kept small but large orientation-dependent non-Gaussianities can be generated.  We study the Suyama-Yamaguchi inequality for the amplitudes of the bispectrum and the trispectrum in the presence of  anisotropic shapes.

\vspace{0.3cm}
%Keywords :  Gravity Waves, Cosmic Strings
\end{abstract}

\date\today

%\preprint{IPM/A-2012/019 }

\maketitle

%%%%%%%%%%%%%%%%%%%%%%%%%%%%%%%%%%%%%%%%%%%%%%%%%%
\section{Introduction}

Recently there have been many interests in anisotropic inflation. This is partially motivated from the cosmological observations indicating some statistical anisotropies on cosmic microwave background (CMB) \cite{Komatsu:2010fb}.  Although the statistical significance of the violation of statistical isotropy is not high \cite{Hanson:2009gu, Hanson:2010gu}, but nonetheless the possibility of having statistically anisotropic seed perturbations are
intriguing.  One useful method to quantify the statistical anisotropy is to write the curvature perturbation power spectrum ${\cal P}_\zeta$ in Fourier space for mode $\vec k$ as \cite{Ackerman:2007nb} ${\cal P}_\zeta = {\cal P}_{0} (1+ g_* (\hat k. \hat n)^2 )$ in which $\hat n$ is the preferred direction in sky. Constraints from CMB and large scale structure indicate that $| g_*| \lesssim 0.4$ \cite{Groeneboom:2009cb, Pullen:2010zy}.

The best method to generate anisotropic perturbations is to employ gauge fields during inflation.  However, due to conformal invariance of $U(1)$ gauge fields in FRW background, the background gauge field energy density and their perturbations will be quickly diluted during inflation. Therefore, one has to break conformal invariance such that the gauge field energy density will not decay during inflation. One popular method is to consider a time-dependent gauge kinetic coupling such that the $U(1)$ action has the form $\Delta {\cal L} = \frac{-f(\phi)^2}{4} F_{\mu \nu} F^{\mu \nu}$ in which $\phi$ is the inflaton field and $F_{\mu \nu}$ is the $U(1)$ gauge field strength. Furthermore, in order for the gauge field perturbations to be scale-invariant one has to choose $f(\phi) \propto a^{-2}$ in which $a(t)$ is the scale factor.
These models in the context of anisotropic inflation and magneto-genesis were studied in great details in literature \cite{Turner:1987bw, Ratra:1991bn, Demozzi:2009fu, Martin:2007ue, Emami:2009vd, Kanno:2009ei, Caldwell:2011ra, Jain:2012ga, Jain:2012vm,  Watanabe:2009ct, Emami:2010rm, Kanno:2010nr, Murata:2011wv, Bhowmick:2011em, Hervik:2011xm, Thorsrud:2012mu, Dimopoulos:2010xq, Yamamoto:2012tq, Moniz:2010cm, Boehmer:2007ut, Koivisto:2008xf, Maleknejad:2011jr, Maleknejad:2012as, Yokoyama:2008xw, Emami:2011yi,   Lyth:2012vn,   Dimopoulos:2009vu, Dimopoulos:2012av,
Dimastrogiovanni:2010sm, ValenzuelaToledo:2009af, Himmetoglu:2008zp, Giovannini:2001nh, Giovannini:2007rh, Kunze:2013hy, Kandus:2010nw, Kahniashvili:2012vt}.

An interesting model of anisotropic inflation was proposed in \cite{Watanabe:2009ct} in which
with $f(\phi) \propto a^{-2}$ the inflationary system admits an attractor solution in which the gauge field energy density, i.e. the electric field energy density, and the metric anisotropy
reaches a small but cosmologically observable level. The cosmological perturbations for this model were studied in \cite{Dulaney:2010sq,  Gumrukcuoglu:2010yc,  Watanabe:2010fh, Yamamoto:2012sq, Funakoshi:2012ym, Bartolo:2012sd}. Similarly, the cosmological  perturbation analysis for a different model of anisotropic inflation \cite{Emami:2011yi} involving a complex inflaton field charged under the $U(1)$ gauge field were performed in
\cite{Emami:2013bk}.  These analysis are based on standard in-in formalism which proved technically difficult due to anisotropic background. On the other hand, experiences with
$\delta N$ formalism \cite{Sasaki:1995aw, Wands:2000dp, Lyth:2004gb, Lyth:2005fi, Naruko:2012fe, Naruko:2012um, Sugiyama:2012tj, Dias:2012qy}
in models of inflation with scalar fields showed that  $\delta N$ analysis are technically
much easier to handle  when calculating the curvature perturbations and their correlations such
as power spectrum and bispectrum. This is specially the case for models in which there are not much interactions when the modes of interest leave the horizon and physically interesting effects, such as non-Gaussianities, originate from local type interactions once the modes are outside the horizon. Therefore it will be very helpful to extend the
standard $\delta N$ formalism \cite{Sasaki:1995aw, Wands:2000dp, Lyth:2004gb, Lyth:2005fi, Naruko:2012fe, Naruko:2012um, Sugiyama:2012tj}
to models of anisotropic backgrounds such as \cite{Watanabe:2009ct}. This is one of our main goal in this work. The applications of  $\delta N$ in models with vector fields were also studied in \cite{Dimopoulos:2008yv, ValenzuelaToledo:2011fj}.

There have been works in the literature employing the conventional
$\delta N$ formalism for the models  containing vector or gauge fields but the effects of
anisotropic background were not taken into account, i.e. the gauge field is treated on the same footing as the scalar fields in an FRW background.
In this work we present a consistent $\delta N$ formalism for anisotropic backgrounds  such as in \cite{Watanabe:2009ct} in which the background metric is in the form of Bianchi I. After presenting our $\delta N$ formalism   we calculate the power spectrum and reproduce exactly the results in \cite{Watanabe:2010fh, Bartolo:2012sd}. We also calculate the bispectrum which coincides exactly with the results of \cite{Bartolo:2012sd}.
%Related to SY inequality we also calculate the trispectrum and discuss the interpretation of
%SY inequality when the shapes of bispectrum and trispectrum are anisotropic.

Planck is expected to release its data soon. Any detection or otherwise of primordial non-Gaussianities from Planck will have significant implications for inflationary model buildings. Simple models of inflation predict almost scale-invariant and almost Gaussian perturbations. Therefore, any detection of primordial non-Gaussianity will go a long way to rule out or classify different inflationary scenarios. Non-Gaussianity may take different shapes in different models, for a review see \cite{Chen:2010xka, Komatsu:2010hc}.  In models of inflation based on scalar fields the shapes of Bispectrum and Trispectrum are statistically isotropic. However, in models of anisotropic inflation, one obtains new shapes which are anisotropic.

As an important consistency condition for single field inflation, a detection of local form bispectrum in the squeezed limit
can rule out all single field models of inflation provided the system reaches the attractor 
solution \cite{Namjoo:2012aa, Chen:2013aj} so one can neglect the evolution of curvature perturbations on super-horizon scales and the curvature perturbations have the initial
Bunch-Davies vacuum state \cite{Agullo:2010ws, Ganc:2011dy}. 
As a different consistency condition, the Suyama-Yamaguchi (SY) inequality \cite{Suyama:2007bg}, \cite{Sugiyama:2011jt}, \cite{Smith:2011if, Assassi:2012zq} between the amplitude of the Bispectrum in the squeezed limit, $f_{NL}$, and the amplitude of the trispectrum in the collapsed limit, $\tau_{NL}$, are expected to hold generally in models of inflation based on scalar fields. It is an interesting question to see if the SY inequality holds when the primordial perturbations are not statistically isotropic.  We will study this question in the context of anisotropic inflation.

The rest of the paper is organized as follows. In Section \ref{deltaN-sec} we present our $\delta N$ formalism. In  Section \ref{anisotropic inflation} we study a model of anisotropic inflation which provides a non-trivial set up to employ our $\delta N$ formalism. In Section \ref{bispectrum} we present the bispectrum and the trispectrum analysis for the anisotropic inflation model and study the
SY inequality. The conclusion and discussions are given in Section \ref{discussions}. We relegates some technical details into Appendices.
\\

{\bf Note added:} While this work was in its final stages the paper \cite{Shiraishi2012} appeared which has overlaps in Bispectrum and Trispectrum analysis in Section \ref{bispectrum} with this work.

%%%%%%%%%%%%%%%%%%%%%%%%%%%%%%%%%%%%%%%%%%%%%%%%%%%%%%
\section{ $\delta N$ formalism for anisotropic backgrounds}
\label{deltaN-sec}

In this section we extend the $\delta N$ formalism \cite{Sasaki:1995aw, Wands:2000dp, Lyth:2004gb, Lyth:2005fi, Naruko:2012fe, Naruko:2012um, Sugiyama:2012tj, Dias:2012qy} to anisotropic backgrounds. 
First we present the background fields equations. After presenting the general metric perturbations,  we look into the fields equations using a gradient expansion method, which is an expansion in $\epsilon$ defined via
\ba
\label{epsilon-eq}
\epsilon \equiv \frac{k}{a H} \, ,
\ea
in which $k$ represents the wave number in Fourier space.  
We demonstrate that the separate universe picture works, that is, in the limit $\epsilon \ll 1$ the background fields equations are locally hold inside each homogenized patch.  This proof is valid to all order in perturbation theory.

%%%%%%%%%%%%%%%%%%%%%%%%%%%%%%%%%%%%%%%%%%%%%%%%%%
\subsection{Background Equations}
Our background is  the Bianchi I metric  with the scale factors $a_1(t), a_2(t)$ and $a_3(t)$ 
\ba
\label{Bianchi-metric1}
ds^2 = -dt^2 + a_1(t)^2 d x^2 +a_2(t)^2 d y^2 +a_3(t)^2 d z^2 \, .
\ea
We adopt the notations used in \cite{Miedema:1993} in which 
\ba
H_i(t)= \dfrac{\dot{a_i}}{a_i} \qquad , \qquad H \equiv \dfrac{1}{3} \sum_{i=1}^3 H_i \, ,
\ea
in which $H_i$ is the Hubble expansion rate for the $i$-th spatial direction, $i=1,2,3$ and a dot indicates the derivative with respect to $t$.

  The components of background Ricci tensor are
\ba
\label{Ricci-back}
R^0{}_0 &=& 3 \dot H + \sum_{k} H_k ^2
\\
R^0{}_i&=&0 
\\
R^i{}_j&=& \delta^i{}_j \left(\dot{H}_i + 3HH_i \right).
%-\delta^i_j \left[\left(\dot{H}_i+3 HH_i \right)+ \dfrac{1}{2} R \right]
\ea
The background Ricci scalar is 
\ba
R =  6 \dot H + 18 H^2 - \sum_{k > k'} H_k H_{k'} \, .
%-3 \dot{H} +9H^2,
%\dfrac{-1}{2} R = \sum_{k} \dot{H}_k +\sum_{k} H^2_k + \sum_{k \neq k'} H_k H_{k'} = \sum_{k} \dfrac{\ddot{a}_k}{a_k} +\sum_{k \neq k'} H_k H_{k'}
\ea
To solve the Einstein fields equations we have to specify our energy momentum tensor.
The general  energy momentum tensor $T_{\mu \nu}$ for an imperfect fluid has the form \cite{ellis98}
\ba
\label{eq:stress}
T_{\mu \nu} = (\rho + p)\,u_{\mu}\,u_{\nu}+ p\,g_{\mu \nu}+ q_{\mu}\,u_{\nu} + u_{\mu}\,q_{\nu} +  \pi_{\mu \nu}
\ea
supplemented with the following conditions 
\ba
 q_{\mu}\,u^{\mu} = 0 \quad , \quad   \pi^{\mu}{}_{\mu} = 0  \quad , \quad  ~\pi_{\mu \nu} = \pi_{\nu \mu} \quad , \quad 
~\pi_{\mu \nu}\,u^{\nu} = 0 \ , \nonumber
\ea
where $u^\mu$ is the fluid's four-vector velocity,  $\rho$ is the relativistic energy density, $p $ is the isotropic pressure, $\pi_{\mu \nu} $ is the trace-free  anisotropic pressure (stress) and $q^\mu$ usually is referred  to as ``heat conduction'', which is also the energy flux relative to $u^\mu$.

The special case of a perfect fluid is identified with $\pi_{\mu \nu}= q^\mu=0$ so we recover the standard form of $T_{\mu \nu}$ for the perfect fluid 
\ba
\label{eq:pf}
T_{\mu \nu} &=& (\rho+p)\,u_{\mu}\,u_{\nu} + p\,g_{\mu \nu}
\quad \quad \mathrm{( perfect \, \,  \,  fluid)} \, .
\ea

For the comoving coordinate associated with the fluid we have 
\ba
u^{\mu} = (1, \vec{0})  \qquad , \qquad u_{\mu} = (-1, \vec{0}),
\ea
so the Einstein equations can be read as 
\ba
\label{00-back}
3  {\cal H}^2&\equiv& \sum_{i > j} \bar H_j \bar H_{j} =  \frac{\bar \rho}{M_P^2}
\\
\bar T^0{}_i &=& \bar q_i =0
\\
\label{i=j-back}
M_P^2 \dot{\bar H}_i&=&-3  M_P^2 \bar H \bar H_i + \dfrac{1}{2} (\bar \rho-\bar p) + \bar \pi^{i}{}_{i}
\ea
Here we have used the convention that $\bar H_i$ represents the background Hubble expansion rates while  $\bar \rho, \bar p$ and  so on represent the background fluid's properties. We also defined
${\cal H}$ as the effective Hubble expansion rate appearing in Friedmann equation, Eq. (\ref{00-back}). Note, ${\cal H}$ here should not be confused with the Hubble expansion rate defined for conformal time usually used in literature.

Finally, the energy conservation equation $u_{\mu} \nabla_{\nu} T^{\mu \nu} = 0$ results in
\ba
\label{cont.-back}
-u_{\mu} \nabla_{\nu} T^{\mu \nu} = \dot{\bar \rho} + 3 H (\bar \rho + \bar p) + \bar H_j \bar \pi
^{i}{}_{j} \delta^{j}_i=0 .
\ea
in which again we have $\bar H = \sum_{i} \bar H_i/3$. 

Note that, in this model, the Hubble parameter appearing in $(0,0)$ component of Einstein equation, ${\cal H}$,  and the Hubble parameter appearing in continuity equation, $\bar H$, are not equal. The  difference between them is given by
\ba
\dfrac{ {\bar H}^2-{\cal H}^2 }{\bar H^2}= \dfrac{1}{6} \, \dfrac{\sum (\bar H-\bar H_i)^2}{\bar H^2}  \, .
%\equiv \dfrac{1}{6} \sigma_{H}^2
\ea
As a result $\bar H > {\cal H}$.

%%%%%%%%%%%%%%%%%%%%%%%%%%%%%%%%%%%%%%%%%%%%%%%
\subsubsection{  Example: $U(1)$ gauge fields in an expanding background}

As an example of non-perfect fluid with anisotropic pressure and heat conduction, consider 
the standard $U(1)$ gauge field theory in an expanding background. This theory will be the base of anisotropic inflation in next section. The action is $L_{em} = -F_{\mu \nu} F^{\mu \nu}/4$ in which $F_{\mu \nu}=\partial_\mu A_\nu - \partial_\nu A_\mu$ is the field strength associated with the $U(1)$ gauge field $A_\mu$.

The electric field, $E_\mu$, and the magnetic field, $H_\mu$,  are given by,
\begin{equation}
E_\mu=F_{\mu\nu}u^\nu
\end{equation}
\\ and
\begin{equation}
H_\rho=\frac 1{2}\eta _{\rho\mu \nu \sigma}u^\mu F^{\nu \sigma} \, .
 \label{mf}
\end{equation}
%where we have
%\begin{eqnarray}
%E_i u^i=0   \quad , \quad  H_i u^i=0  \end{eqnarray}
The electromagnetic energy-momentum tensor, $T^{\mu \nu}_{em}$, is
\begin{equation}
T^{\mu \nu}_{em}= F^{\sigma \mu}F_\sigma^{\hspace{1mm}\nu}- \frac{1}{4}g^{\mu \nu}F_{\sigma \rho}F^{\sigma \rho} \, .
\label{Tem}
\end{equation}
For  an observer comoving with the fluid $T_{em}^{\mu \nu}$ can be written as \cite{barrow97} 
\begin{equation}
T_{em}^{\mu \nu}=\frac12\left( E^2+H^2\right) u^\mu u^\nu+\frac
16\left( E^2+H^2\right) h^{\mu \nu}+2u^{(\mu}\eta
^{\nu)\alpha \beta \gamma}u_\alpha E_\beta H_\gamma+\pi^{\mu \nu},  \label{Tem1}
\end{equation}
where $\eta^{\mu \nu \alpha \beta}$ is the four-dimensional totally antisymmetric volume element  ($\eta_{0123} = \sqrt{- \det g}$), $h_{\mu \nu} = g_{\mu \nu} + u_\mu u_\nu$ is the projection matrix,  $E^2=E_\mu E^\mu$ and $H^2=H_\mu H^\mu$, respectively, are the magnitudes of the  electric and the magnetic fields and $\pi_{\mu \nu}$ is a traceless and  space-like symmetric tensor given by
\begin{equation}
\pi^{\mu \nu}_{em}= \frac{1}{3}\left(E^2+H^2\right)h^{\mu \nu}- E^\mu E^\nu-H^\mu H^\nu .  \label{Mten1}
\end{equation}
Eq. (\ref{Tem1}) can be compared with the energy momentum tensor for a generic
imperfect fluid defined in Eq. (\ref{eq:stress}) which yields 
\begin{eqnarray}
\rho _{em} &=&\frac 1{2}\left( E^2+H^2\right) ,  \label{muem}
\\
&&  \nonumber \\
p_{em} &=&\frac 16\left( E^2+H^2\right) ,  \label{pem} \\
&&  \nonumber \\
q_{em}^\mu &=&\eta ^{\mu \nu   \alpha \beta}u_\nu E_\alpha H_\beta,  \label{qem} \\
&&  \nonumber \\
\pi ^{\mu \nu } &=&\pi^{\mu \nu}_{em}.  \label{piem}
\end{eqnarray}

%%%%%%%%%%%%%%%%%%%%%%%%%%%%%%%%%%%%%%%%%%%%%%%%%%
\subsection{Perturbations}

Let us now consider the fields equations with perturbations. In our $\delta N$ analysis we adopt the notation used in \cite{Sugiyama:2012tj}. The order of spatial derivative or the so-called gradient expansion is denoted by $\epsilon=k/aH$ while the order of smallness of perturbations are denoted by $\delta$. In principle, one has to consider different gradient expansion parameters $\epsilon_i$ for different directions $\epsilon_i = k/a_i H_i$. However, to simplify the analysis we assume $\epsilon_i \sim \epsilon$ so there is no hierarchy for gradient expansions along different directions. 

We use the standard ADM formalism for the metric decomposition as follows
\ba
\label{ADM}
ds^2 = -d{\cal N}^2 + \gamma_{ij} \left( dx^i + \beta^i dt \right) \left( dx^j + \beta^j dt\right) \, ,
\ea
in which ${\cal N}$ is the lapse function, $\beta_i$ are the shift vectors, and $\gamma_{ij}$ represent the spatial three-dimensional  metric. The spatial indices $i=1,2,3$ are raised or lowered by the spatial metric  $\gamma_{ij}$. Furthermore,  we decompose the spatial metric as follows
\ba
\label{gamma-ij}
\gamma_{ij} = a_i(t) a_j(t) e^{\psi_i (\mathbf{x},t)+\psi_j (\mathbf{x},t)} \tilde{\gamma}_{ij} \, ,
\ea
where $a_i(t)$ is the average scale factor for the $i$-th spatial direction and $\psi_i(\mathbf{x},t)$  are equivalent to curvature perturbation $\psi$ in the isotropic limit. In linear perturbation theory $ \beta^i, \psi_i $ and $ \tilde {\gamma}_{ij}$ are small perturbations at the order  ${\cal O}({\delta})$ with $\delta \ll1$.  But in our analysis below, we  do not use the assumption that
$\delta \ll 1$ so our analysis are valid to all order in perturbation theory. 

Note that in general Bianchi Type-I model we considered here  there is no spatial symmetry so all physical degrees of freedom are in the form of scalar perturbations, encoded  in ${\cal N}, \beta_i, \psi_i$ and $\tilde{\gamma}_{ij} , i \neq j$
and  there is no vector or tensor perturbations.

%In matrix notation, our general metric has the following form 
%\ba
%\label{deltag}
% g_{\alpha \beta} = \left(
%\begin{array}{c}
%- {\cal N}^2 + \beta_i \beta^i~~~~~~~~ a^2_1e^{2\psi_1} \beta^1~~~~~~~~~~~a^2_2e^{2\psi_2} \beta^2~~~~~~~~~~~~~~a^2_3 e^{2\psi_3} \beta^3
%\\\\
%~~~~~~~~~~~~~~~~~~~~~~~  a_1^2 e^{2\psi_1} ~~~~~~~~~~~~~~a_1a_2\tilde{\gamma}_{12}~~~~~~~~~~~~~~~~a_1a_3\tilde{\gamma}_{13}
%\\\\
%~~~~~~~~~~~~~~~~~~~~~~~~~~~~~~~~~~~~~~~~~~~~~  a_2^2 e^{2 \psi_2}~~~~~~~~~~~~~~~~~a_2a_3\tilde{\gamma}_{23}
%\\\\
%~~~~~~~~~~~~~~~~~~~~~~~~~~~~~~~~~~~~~~~~~~~~~~~~~~~~~~~~~~~~~~~~~~~~~~~~ a_3^2 e^{2\psi_3}
%\end{array}
%\right ) \hspace{0.5cm}
%\ea

An important step in dealing with the gradient expansion ordering of Einstein equations 
is the order of the shift functions $\beta^i$. We note that at the background level 
$\beta^i=0$. As a result one expects that the background metric should be valid globally 
in the limit $\epsilon \rightarrow 0$ and, as employed in  \cite{Lyth:2004gb}, one can assume 
\ba
\label{beta-order}
\beta^{i} = {\cal O} (\epsilon).
\ea
The ordering of $\beta^i $  in Eq. (\ref{beta-order}) was also obtained in \cite{Sugiyama:2012tj} with the assumption that the anisotropic pressure is first order in gradient expansion. 
We look into ordering of $\beta^i$ more rigorously in Appendix \ref{App-B}  and verify Eq. (\ref{beta-order}).  
Furthermore, as we demonstrated in Appendix  \ref{off-Ein}, 
it can be shown that the non-diagonal spatial metric components, $\gamma_{ij}$, to all orders in perturbations  theory  are also at the first order of gradient expansion
\ba
\gamma_{i\neq j} = {\cal O} (\epsilon) \, .
\ea
% The detailed discuusion can be found in Appendix  \ref{off-Ein}

Now we have all the necessary materials for performing the gradient expansion analysis for the Einstein equations. Here we emphasis that  the following expansions are valid to the first order of gradient expansion $\epsilon$ but to all orders of perturbations $\delta$.

The $(0,0)$ component of perturbed Einstein tensor is
\ba
\label{00-pert}
G^0{}_0 =  \dfrac{-1}{{\cal N}^2}  \sum_{i > j} (\bar H_i+ \dot{\psi}_{i})(\bar H_j+ \dot{\psi}_{j}) +{\cal O} (\epsilon^2)
\ea
%in which $\bar{\cal H}$ is defined in background Friedmann equation, Eq. (\ref{00-back}).
Combining Eq. (\ref{00-pert}) with the background $(0,0)$ component 
equation, Eq. (\ref{00-back}), yields 
 \be
 \dfrac{-M_P^2}{{\cal N}^2}  \sum_{i > j} (\bar H_i+ \dot{\psi}_{i})(\bar H_j+ \dot{\psi}_{j})=  \rho (\mathbf{x},t)+O(\epsilon^2)
\ee
Locally, as a function of $(\mathbf{x},t)$, the above equation takes the form
\be
\label{Fried-pert}
3 M_P^2 {\cal H}^2(\mathbf{x},t) =  \rho (\mathbf{x},t) +O(\epsilon^2),
\ee
in which
\ba
\label{calH}
 {\cal H}^2(\mathbf{x},t) \equiv \frac{1}{3}\sum_{i > j} H_i(\mathbf{x},t) H_{j} (\mathbf{x},t) \, ,
\ea
with the following generalization of local Hubble expansion parameter $H_i(\mathbf{x},t)$
\ba
\label{H-local}
H_i(\mathbf{x},t) \equiv \dfrac{ \bar H_i(t) + \dot{\psi}_i (\mathbf{x},t)}{\cal N} \, .
\ea
As a result one can readily associate the average local Hubble expansion rate $H (\mathbf{x},t) $ as 
\ba
\label{H-average}
 H (\mathbf{x},t) \equiv  \frac{1}{3} \sum_i H_i(\mathbf{x},t) = \dfrac{ \bar H(t) + \frac{1}{3}\sum_i\dot{\psi}_i (\mathbf{x},t)}{\cal N}
\ea
in which the background average Hubble expansion rate $\bar H$ is $\bar H= \sum_i \bar H_i/3$.

Now we look at the energy conservation equation in its contracted form 
$u_{\mu} \nabla_{\nu} T^{\mu \nu} = 0$.  At the background level the energy conservation
equation is given by Eq. (\ref{cont.-back}). Defining the fluid's proper time $\tau$ via
$\frac{d}{d\tau} = u^{\mu} \nabla_{\mu} \simeq \frac{1}{\cal N} \frac{d}{dt} +{\cal O}(\epsilon^2)$, the perturbed
energy conservation equation is
\ba
%\label{cont-eq-1}
\dfrac{d\rho(\mathbf{x},t)}{d \tau}+3 H(\mathbf{x},t) \left(\rho(\mathbf{x},t)+p(\mathbf{x},t)\right) +\left[-u_{\mu} \frac{d}{d\tau}q^{\mu} + \nabla_{\mu} q^{\mu}- u_{\mu} \nabla_{\nu} \pi^{\mu \nu}\right]=O(\epsilon^2)\,,
\label{cont}
\ea
in which $H(\mathbf{x},t)$ is the average local Hubble expansion rate defined in Eq. \eqref{H-average}. By using Eq. \eqref{aniso-pres-con} and \eqref{heat-con} the above equation takes the following simple local form
\ba
\dfrac{d\rho(\mathbf{x},t)}{d \tau}+3 H(\mathbf{x},t) \left( \, \rho (\mathbf{x},t)+p(\mathbf{x},t) \, \right)+ \sum_i{\pi}^i{}_i(\mathbf{x},t)   H_i (\mathbf{x},t) ={\cal O}(\epsilon^2) \, .
\label{cont-sim}
\ea
in which ${\pi}^i{}_i(\mathbf{x},t) =  \bar {\pi}^i{}_i + \delta {\pi}^i{}_i(\mathbf{x},t)$  to all orders in perturbations.

So again we conclude that  our separate universe  recipe works and it is enough to replace any background function $f(t)$ by its local form $f(\mathbf{x},t)$ and also using new local directional Hubble parameters $H_i (\mathbf{x},t)$. Our prescription will be satisfactory if we can also check the $(i=j)$ components of Einstein equations which are identical to the dynamical equations  of $\pi^i_i$. The diagonal spatial components of Ricci tensor  can be read as 
%\ba
%R^i{}_i =(\dot{\bar H}_i + 3\bar H \bar H_i )(1-2A) + \ddot{\psi}_i + 3 \bar H \dot{\psi}_i  + \bar H_i \sum_{j} \dot{\psi}_j - \bar H_i \dot{A} + {\cal O} (\epsilon^2,\delta^2) \, .
%\ea
%One can simply show that the above equation can be simplified as follows
\ba
R^i{}_i =\dfrac{dH_i(\mathbf{x},t)}{d \tau} + 3 H(\mathbf{x},t)H_i(\mathbf{x},t) + {\cal O} (\epsilon^2) \, ,
\ea
so the $(i=j)$ components of Einstein equation simply modifies the corresponding background equation, Eq. (\ref{i=j-back}), as follows (for the off-diagonal components of Einstein equation see Appendix \ref{off-Ein})
\ba
\label{dH-pert}
M_P^2 \dfrac{dH_i(\mathbf{x},t)}{d \tau}&=&-3 M_P^2 H(\mathbf{x},t)H_i(\mathbf{x},t) + \dfrac{1}{2} \left(\rho(\mathbf{x},t)-p(\mathbf{x},t) \right)  + \pi^{i}{}_{i}(\mathbf{x},t)
\ea

Now we have a complete set of local fields equations, Eq. (\ref{Fried-pert}), Eq. (\ref{cont-sim}) and Eq. (\ref{dH-pert}),  mimicking the corresponding background equations,
Eq. (\ref{00-back}),  Eq. (\ref{cont.-back}) and Eq. (\ref{i=j-back}),  
with the  local Hubble parameters $H_i(\mathbf{x},t) $ defined in Eq. (\ref{H-local}).  We emphasize again that this set of equations are valid to all order in perturbations $\delta$ but to the first order of gradient expansion $\epsilon$.

The separate Universe approach discussion is now complete. The $\delta N$ formalism is also at hand noting that from the  equations above one has
\ba
N_i(\mathbf{x},t_1,t_2) \equiv \int_{t_1}^{t_2} H_i (\mathbf{x},t) {\cal N} dt = \int_{t_1}^{t_2} \bar H_i  dt + \int_{t_1}^{t_2} \dot{\psi}_i  dt
\ea
So one readily finds
\ba
N_i(\mathbf{x},t_1,t_2) - \bar{N}_i(t) = \psi_i(t_2) -  \psi_i(t_1)
\ea
Now defining the average  expansion by 
\ba
N(\mathbf{x},t_1,t_2)= \frac{1}{3}\sum_i N_i(\mathbf{x},t_1,t_2) 
= \int_{t_1}^{t_2} H (\mathbf{x},t) {\cal N} dt
\ea
 one obtains
\ba
\delta N(\mathbf{x},t_1,t_2)=      N(\mathbf{x},t_1,t_2) - \bar{N}(t) = \psi(t_2) -  \psi(t_1)
\ea
in which   $\psi (\mathbf{x},t)$ is defined as the average of $\psi$
\ba
\psi(\mathbf{x},t) \equiv \dfrac{1}{3}  \sum_{i} \psi_i(\mathbf{x},t) \, .
\ea

We are interested in curvature perturbation on surface of constant energy density. As was demonstrated in Appendix \ref{gauge-transformations}, the curvature perturbation $\zeta$ defined in  Eq. (\ref{zeta-def}) via
\ba
\label{zeta}
-\zeta  = \psi - \frac{H}{\dot \rho} \delta \rho \, ,
\ea
is gauge invariant.  But this definition just works to the first order in perturbations $\delta$. The definition of $\zeta$
to all orders of perturbation theory can be found in \cite{Lyth:2004gb}. However, as it is shown below, we calculate  $\delta N$ on the surface of uniform energy density so the definition of $\zeta$ to nonlinear orders is irrelevant for our purpose. 

The relation between $\zeta$ and $\delta N$ therefore is  
\ba
\label{zeta-N}
\zeta(\mathbf{x},t) = \delta N(\mathbf{x},t_i,t_f)  \, ,
\ea
in which the initial surface is a flat surface $\psi=0$ and the final surface should be a uniform energy density surface $\delta \rho= 0$.

Here a comment is in order. The  diagonal components of the anisotropic pressure,  $\delta \pi^{i}_i$ (no sum over $i$), are non-zero at the background level  so their perturbations are expected to play some roles in the curvature perturbation analysis.  However, the non-diagonal spatial components of anisotropic pressure and the heat conduction terms are absent at the background level so their perturbations will dilute quickly. The diagonal anisotropic pressure plays two different roles in the curvature perturbation analysis, a direct effect and an indirect effect. The direct effect can be seen from the continuity equation, Eq. \eqref{cont-sim}, in which $\delta \pi^{i}_i$ contributes to the Hubble expansion rate. This effect, by using Eq.\eqref{cont-sim}, can be quantified as follows
 \ba
 \label{N-formula}
N(\mathbf{x},t_i,t_f) =\int_{t_i}^{t_f} H(\mathbf{x},t)   d\tau=  -\dfrac{1}{3} \int_{t_i}^{t_f}  dt \dfrac{\dot{\rho}(\mathbf{x},t)}{\rho(\mathbf{x},t)+p(\mathbf{x},t)} - \dfrac{1}{3} \int_{t_i}^{t_f} dt Q(\mathbf{x},t)  \, ,
\ea
in which $Q(\mathbf{x},t) $ is defined as
\ba
\label{Q-eq}
Q(\mathbf{x},t) &=& \dfrac{\cal N}{\rho+p} \left[-u_{\mu} \frac{d}{d\tau}q^{\mu} + \nabla_{\mu} q^{\mu}- u_{\mu} \nabla_{\nu} \pi^{\mu \nu}\right] \nonumber\\
& =& \dfrac{{\cal N}(\mathbf{x},t)}{\rho(\mathbf{x},t)+p(\mathbf{x},t)} \sum_i H_i(\mathbf{x},t) \pi^i{}_i (\mathbf{x},t)  + {\cal O} (\epsilon^2).
\ea
The above equation shows that the diagonal anisotropic pressure components $\delta \pi^{i}{}_{i}$ contributes to $\delta N$ through their effect on continuity equation as captured by
the term containing $Q$ in Eq. (\ref{N-formula}).

The  indirect effect of anisotropic pressure is more subtle and sometimes can be more important than the contribution from the term containing $Q$ above. This effect can be understood as the back-reactions of fields responsible for anisotropic pressure on the dynamics of other background fields  such as the inflaton field. The $\delta N$ formalism automatically includes this indirect effect. We will see this effect in next section in application of our $\delta N$ formalism for models of  anisotropic inflation.

%%%%%%%%%%%%%%%%%%%%%%%%%%%%%%%%%%%
\section{Anisotropic Inflation}
%%%%%%%%%%%%%%%%%%%%%%%%%%%%%%%%%
\label{anisotropic inflation}

In this section we present the model of anisotropic inflation with a $U(1)$ gauge field originally presented in \cite{Watanabe:2010fh} which provide a non-trivial setup to employ our
$\delta N$ formalism.

The action is given by
\ba
\label{action}
S= \int d^4 x \sqrt{-g} \left [ \frac{M_P^2}{2}
R -  \frac{1}{2} \partial_\mu \phi\partial^\mu \phi - \frac{f^2(\phi)}{4}F_{\mu \nu} F^{\mu \nu}  - V(\phi) \right]
\ea
in which $\phi$ is the inflaton field and $F_{\mu\nu} = \partial_\mu A_\nu - \partial_\nu A_\mu $
is the field strength associated with the $U(1)$ gauge field $A_\mu$.

To employ the $\delta N$ formalism, as usual we need to have a good control of the background dynamics.  We assume that the gauge field has a non-zero classical value along the x-direction so $A_\mu =(0,A_{x}(t),0,0)$. As a result, the background space-time is in the form of  Bianchi I Universe with the metric 
\ba
%\label{Bianchi1}
ds^2 &=& - dt^2 + e^{2\alpha(t)}\left( e^{-4\sigma(t)}d x^2 +e^{2\sigma(t)}(d y^2 +d z^2) \right)  \nonumber\\
&=&   - dt^2 + a(t)^2 dx^2 + b(t)^2 (dy^2 + dz^2)
\ea
In this view $H \equiv \dot \alpha$ is the average Hubble expansion rate, $H_a \equiv \dot a/a$
and $H_b \equiv \dot b/b$ are the expansion rates along the spatial directions $x$ and $y$  
and $\dot \sigma/H \equiv (H_b - H_a)/ H$ is a measure of anisotropic expansion.

%%%%%%%%%%%%%%%%%%%%%%%%%%%%%%%%%%%%%%%%%%%%%%%%%%
\subsection{The background dynamics}

The fields equations are given by
\ba
\label{back-A-eq}
\partial_t{\left(  f^2(\phi) e^{\alpha + 4 \sigma} \dot A_x        \right)}& =& 0 \\
%\ddotA_{x}+\left[\dot\alpha+4\dot\sigma+2\frac{f_\rho(\rho)}{f(\rho)}\dot \rho
%\right]\dot A_{x}+e^2\rho^2f^{-2}(\rho)A_{x}&=&0 \, , \\
\label{back-rho-eq}
\ddot\phi+3\dot \alpha\dot \phi+ V_\phi
-f(\phi)f_{,\phi}(\phi)\dot A_x^2   e^{-2\alpha+4\sigma}&=&0  \\
\label{Ein1-eq}
\frac{1}{2}\dot
\phi^2+V(\phi)+    \frac{1}{2}f^2(\phi)\dot
A_x^2  e^{-2\alpha+4\sigma}
&=&
3 M_P^2 \left(   \dot \alpha^2-\dot \sigma^2 \right)  \\
\label{Ein2-eq}
V(\phi)+   \frac{1}{6}f^2(\phi)\dot
A_x^2  e^{-2\alpha+4\sigma}
&=& M_P^2 \left( \ddot \alpha    + 3 \dot \alpha^2 \right)  \\
\label{anisotropy-eq}
\frac{1}{3}f^2(\phi)\dot A_x^2   e^{-2\alpha+4\sigma}
&=& M_P^2\left( 3\dot \alpha \dot \sigma+ \ddot \sigma      \right)\, ,
\ea
in which a dot indicates derivative with respect to $t$.

The equation of motion for $A_{x}$ (the Maxwell equation) is easily solved as
\ba
\label{gaugefield}
\dot{A_{x}}= f(\phi)^{-2}e^{-\alpha(t)-4\sigma(t)}p_{A} \, ,
\ea
where $p_{A}$ is a constant of integration.

We are interested in the small anisotropy limit, $|\dot \sigma/H | \ll 1 $, so the background expansion is mainly supported by the isotropic potential term as in conventional models of inflation.  In order
for the anisotropy to be small, we demand that $R \ll 1$ in which
\ba
\label{R-def}
R \equiv \frac{\dot A^2 f(\phi)^2 e^{-2 \alpha}}{2 V} \, .
\ea
In this view $R$ measures the ratio of the electric field energy density, $\rho_{em}$, associated with the gauge field  to the total potential energy density. Therefore, to have  small anisotropies, we require $\rho_{em} \ll V$.

Although the anisotropy is small, $R\ll 1$, so the Hubble expansion rate in modified Friedmann equation (\ref{Ein1-eq}) is mainly dominated by the isotropic potential term, but the back-reactions of the gauge field on the inflaton field induce an effective mass for the inflaton  as given by the last term in Eq. (\ref{back-rho-eq}). This in turn will affect the dynamics of the inflaton field. As shown in \cite{Watanabe:2009ct} with the appropriate form of $f(\phi)$ the
the system reaches an attractor solution in which $R$ reaches a subdominant but nearly constant value. For $R$ to be constant, we need $f(\phi) \propto a^{n}$
with $n\simeq -2$. Indeed, the background expansion is given by
\ba
\label{a-scale}
a \propto \exp \left[ - \int d \phi \frac{V}{ V_\phi} \right] \, .
\ea
So if one chooses
\ba
\label{f-scale}
f \propto \exp \left[ -n  \int d \phi \frac{V}{V_\phi} \right]
\ea
this yields $f \propto a^{n}$. The exact form of $f$ therefore depends on $V(\phi)$. For the chaotic potential used in \cite{Watanabe:2009ct}  we have
\ba
\label{Chaoticpotential}
V= \frac{1}{2} m^2 \phi^2  \quad \quad  \rightarrow \quad \quad
f(\phi) = \exp {\left( \frac{c\phi^2}{2 M_P^2}  \right)}
\ea
with $c$ a constant very close to unity.  In our discussion below we take the form of $f$, in
terms of $a= e^{\alpha}$, to be
\ba
\label{f-form}
f= \left( \frac{a}{a_f} \right)^{-2 c}  \simeq \left( \frac{\eta}{\eta_e}\right)^{2c} \, ,
\ea
in which $a_e$ and $\eta_e$ represent the value of the scale factor and the conformal time at the end of inflation.

As shown in \cite{Watanabe:2009ct} the system reaches the attractor solution in which
$R$ is given by
\ba
\label{R-app}
R = \frac{c-1}{2c}\epsilon_{H} = \frac{1}{2}I\epsilon_{H} \, ,
\ea
where we have defined $I\equiv\frac{c-1}{c}$ and $\epsilon_H \equiv \dot H/H$ is the slow-roll parameter. Combined with the definition of $R$ in Eq. (\ref{R-def}) we obtain
\ba
\label{cons-eq}
\dot A^2 f^2 e^{-2 \alpha} = I \, \epsilon_H V \, .
\ea
As we shall see below, this equation will be the key equation to find $\delta N$ in terms of $\delta \phi$ and $\delta \dot A$.

Furthermore the anisotropy in expansion is given by
\ba
\label{anis-H}
\frac{\dot \sigma}{\dot \alpha} \simeq \frac{I \epsilon_H}{3}\, .
\ea
During the attractor phase the inflaton evolution is given by
\ba
\label{klinGordon}
M_P^{-2}\frac{d \phi}{d \alpha} \simeq -\frac{V_\phi}{ V} + \frac{c-1}{c}\frac{V_\phi}{ V} \, .
\ea
Interestingly, this means that the back-reactions of the gauge field on the inflation field change the effective mass of the inflaton field as given by the second term above.

Using Eq. (\ref{Chaoticpotential}) in Eq. (\ref{klinGordon}) results in the following equation
\ba
\label{klinN}
\phi_{e}^2 - \phi^2 = 4 M_P^2 \alpha (1-I)
\ea
in which $\phi_e$ is the value of $\phi$ at the end of inflation. We choose the convention such that $\alpha_e=0$, so during inflation $\alpha<0$. Eq. (\ref{klinN}) clearly shows the effect of the gauge field back-reactions on the evolution of the inflaton field. The fact that the
evolution of the inflaton field is affected by the gauge field, as given by the correction factor $(1- I)$ in Eq. (\ref{klinN}) is the key to calculate $\delta N$ in the presence of gauge field. In passing, we comment that in the previous applications of $\delta N$ in the literature for models with the  gauge fields, this important effect is not taken into account. In other words, $\delta N$ in these papers  have been written with treating $\delta A_\mu$ in the same footing as $\delta \phi$ in  an FRW background without taking into account the back-reactions  of the gauge field
in the evolution of inflaton field and in the dynamics of the anisotropic background.

In connection with our discussion in previous section the energy density, pressure, momentum density   and  stress associated with the electro-magnetic field are given by
\begin{eqnarray}
\label{muem-eq}
\rho _{em} &=&\frac 1{2}\left( E^2+B^2\right) = \frac{3}{2}I \epsilon_{H}H^2 ,  \label{muA}\\
p_{em} &=&\frac 16\left( E^2+B^2\right) = \frac{1}{2}I \epsilon_{H}H^2  ,  \label{pA} \\
q_{em}^i &=&\eta ^{ijkq}u_jE_kB_q =0 ,  \label{qA} \\
\end{eqnarray}
and
\be
\pi_{\mu}^{\nu}=\left[ \begin{array}{cccc}
0&0&0&0\\
0&-2I\epsilon_{H}H^2&0&0\\
0&0&I\epsilon_{H}H^2&0\\
0&0&0&I\epsilon_{H}H^2
\end{array}\right ].
\ee
Plugging the value of $\rho_{em}$ into definition of $R$ and using the attractor value
Eq. (\ref{R-app}) we obtains $\rho_{em} \simeq R V$ as advertised before. Also Eq. (\ref{muem-eq}) indicates that  $E=\sqrt{3I\epsilon_{H}}H$ and $B=0$.

%%%%%%%%%%%%%%%%%%%%%%%%%%%%%%%%%%%%%%%%%%%%%%%%%%

\subsection{$\delta N$ in anisotropic inflation}

Our goal here is to calculate the curvature perturbations in this model by employing our
$\delta N$ formalism. As we argued before the contribution of the gauge field into the Hubble expansion rate and total energy density is sub-dominant. This means that the surface of end of inflation is controlled only by the inflaton field. However, the gauge field plays an important role in Klein-Gordon equation and in the evolution of the inflaton field as can be seen in Eq.(\ref{klinN}). 

Perturbing  Eq.(\ref{klinN}) we have,
\ba
\label{pert-klin}
2\phi \delta \phi = -4\delta N + 4N \delta I \, .
\ea
As a result
\ba
\label{delta N}
\delta N = -\frac{1}{2}\phi \delta \phi + N\delta I \, .
\ea
Note that, in order to connect to the standard notation we made the replacement $\alpha \rightarrow N$ so $\delta \alpha = \delta N$ from now on. Also note that $R$ is related to $I$ by Eq. (\ref{R-app}) so by $\delta I$, we actually mean $\delta (R/\epsilon)$.
Since it is easier to work with $\delta I$ than $\delta R$, we use $\delta I$ from now on.

The first term in Eq. (\ref{delta N}) is the contribution of the inflaton field, while the second term is due to the  back-reaction of the gauge field on the inflaton dynamics.    Also by perturbing Eq. (\ref{cons-eq}) we have
\ba
\label{deltaA1}
\epsilon_{H} \delta I = -12 R \delta N + 4 R \frac{\delta \dot{A_x}}{\dot{A_x}}
\ea
Now combining Eq. (\ref{delta N}) and Eq. (\ref{deltaA1}) we have
\ba
\label{total delta N}
\left(1 + \frac{12 R N}{\epsilon_{H}}\right)\delta N = -\frac{\phi}{2 M_P^2} \delta\phi + \frac{4R N }{\epsilon_{H}}\frac{\delta \dot{A_x}}{\dot{A_x}}
\ea
Now using Eq. (\ref{R-app}) we have $R N/\epsilon_H \sim I N$. As we shall see, we require $N I  \ll 1$ in order not to produce too much anisotropy in power spectrum so we can neglect the second term in the left hand side of Eq. (\ref{total delta N}) and
\ba
\label{App delta N}
\delta N \simeq -\frac{\phi}{2 M_P^2} \delta\phi + 2 I N \frac{\delta \dot{A_x}}{\dot{A}} \, .
\ea
This is our result for $\delta N$ to linear order in terms of $\delta \phi$ and $\delta \dot A$. Interestingly, since in this model the leading contribution into the anisotropic power spectrum
comes from the electric field instead of the magnetic field, we see that in $\delta N$ only 
$\delta \dot{A}$ and not $\delta A$ appears. This should be compared with the conventional models of $\delta N$ involving scalar fields $\phi_I$ in which
$\delta \phi_I$ and not $\partial_t  \delta {\phi_I}$ appears. This is because for light scalar fields $\delta \dot \phi_I$ become negligible on super-horizon scales once the attractor solution has been reached. 

To calculate the power spectrum and the higher order correlations, we have to know the behavior of  $\frac{\delta \dot{A_x}}{\dot{A}}$ outside the horizon. For this purpose, we have to solve the mode function for $\delta A_i$ with the initial Bunch-Davies vacuum deep inside the horizon.  As shown in \cite{Bartolo:2012sd} the canonically normalized gauge field quantum fluctuations are given by 
\ba
\label{canonical gauge field}
\delta A_i =   \sum_{\lambda = \pm} \int \frac{d^3k}{(2\pi)^{3/2}}e^{i \overrightarrow{k}.\overrightarrow{x}}\vec{\epsilon}_{\lambda}(k) \frac{\widehat{V}_i}{f}  \, ,
\ea
in which
\ba
\widehat{V} = a_{\lambda}(\overrightarrow{k})V_{\lambda}(k) + a_{\lambda}^{\dagger}(-\overrightarrow{k})V_{\lambda}^{*}(k) \, .
\ea
Here  $a_{\lambda}(\overrightarrow{k})$ and $a_{\lambda}^{\dagger}(\overrightarrow{k})$
represent the annihilation and the creation operators and $\epsilon_\lambda$ for $\lambda =\pm$ represents the circular polarization with the properties
$\vec k \,  . \,   \vec \epsilon_\pm(\vec k) =0 \, , \,  \vec k \, \times \,  \vec \epsilon_\pm(\vec k) =
\mp i k \, \vec \epsilon_\pm(\vec k)  \, , \vec \epsilon_\lambda (-\vec k) = \vec \epsilon_\lambda\, (\vec k)^*$, normalized via
$\vec \epsilon_\lambda(\vec k)  \, . \,  \vec \epsilon_{\lambda'}(\vec k')= \delta_{\lambda \lambda'}$ and
\ba
\sum_{\lambda}\epsilon_{\lambda,i}(\vec{k})\epsilon^{*}_{\lambda,j}(\vec{k}) = \delta_{ij} - \hat{k}_{i}\hat{k}_{j} \, .
\ea

The mode functions satisfy the evolution equation
\begin{align}
\label{canonical gauge field equation}
V_{\lambda}(k)''+ \left( k^2 - \frac{f''}{f} \right)V_{\lambda}(k)=0 \, ,
\end{align}
where the prime denotes the derivative with respect to conformal time $d \eta= dt/a(t)$. For  $f$ given in 
Eq. (\ref{f-form}) the normalized gauge field mode function is the same as that of  a massless scalar field in dS space with
\begin{align}
\label{canonical gauge field}
V_{\lambda}(k) \simeq \frac{1+ i k \eta}{\sqrt{2}k^{3/2}\eta} e^{-ik\eta} \, .
\end{align}
Using this form of the wave function and the attractor solution Eq. (\ref{cons-eq})
one can easily show that on super-horizon scales
\ba
\label{mode-deltaA}
\frac{\delta \vec{\dot{A}}}{\dot{A}} = \sum_{\lambda}\vec{\epsilon}_{\lambda} \frac{\sqrt{3}H}{\sqrt{2I\epsilon_{H}k^3}}    \quad \quad (k > a H)
\ea
 In particular, we see that on super-horizon scale, $ \delta A_x/\dot A_x$ is a constant.

Now we are in the position to calculate the total effect of the gauge field in curvature perturbation $\zeta$.  From Eq. (\ref{N-formula})  and Eq. (\ref{App delta N})
we see that there are two different terms that encode the contributions of the gauge field in $\zeta$.  Eq. (\ref{App delta N}) encodes the indirect effects of the gauge field on $\delta N$ originating from its back-reaction on inflaton field dynamics. However, the direct
contribution of the gauge field in $\delta N$ is encoded in term $Q$ in  Eq. (\ref{N-formula}).
As we shall prove below, the contribution from the $Q$ term in the curvature perturbation 
is negligible and the leading contribution of the gauge field in curvature perturbation is 
from its back-reaction effects in Eq. (\ref{App delta N}).

Calculating $Q$ from Eq. (\ref{Q-eq}) yields (note that in this model $q^{\mu}$ is proportional to the product of the electric and magnetic fields and since in this model the magnetic field is zero therefore there is no correction from $q^{\mu}$)
\begin{eqnarray}
\label{Q app term}
Q &=& \frac{1}{(\rho+p)}\left( H_{a}\pi^{1}_1 + 2 H_{b}\pi^{2}_2 \right)\nonumber\\
~&=& \frac{2H^2}{(\rho+p)}\left(H_{b} - H_{a} \right)I \epsilon_{H}\nonumber\\
~&=& I^2 H \epsilon_{H} \, ,
\end{eqnarray}
where we have used $(\rho+p) \simeq \dot{\phi}^2 = 2 H^2 \epsilon_{H}$ and $H_{b} - H_{a}= H I\epsilon_{H}$ from Eq. (\ref{anis-H}).

Perturbing Eq. (\ref{Q app term}), we have
\begin{eqnarray}
\label{Q pert term}
\delta Q &=&2 I H\epsilon_{H}\delta I \nonumber\\
~&=& 2 I H \left(-12 R \delta N + 4 R \frac{\delta \dot{A_x}}{\dot{A}}\right) \, .
\end{eqnarray}
Now integrating Eq. (\ref{Q pert term}) over $t$ we have,
\begin{eqnarray}
\label{int Q pert term}
\int_{t_{1}}^{t_{2}}\delta Q  dt&=& \int_{t_{1}}^{t_{2}}  2 H I\left(-12 R \delta N + 4 R \frac{\delta \dot{A_x}}{\dot{A}}\right) d t \nonumber\\
~&=&  2 I N\left(-12 R \delta N + 4 R \frac{\delta \dot{A_x}}{\dot{A}}\right)
\end{eqnarray}
To perform this integral, we have assumed hat $H$ and $R$ are nearly constant in the slow-roll approximation. Furthermore, $\delta \dot A_x/\dot A_x$ is also nearly constant as can be seen from  Eq. (\ref{mode-deltaA}).

Eq. (\ref{int Q pert term}) indicates that the contribution of $Q$ in $\delta N$ is at
the order of $I N R \sim N I^2$. However, as we shall see, in order not to produce too much
anisotropy we require $I N^2 < 1$ so $ N I^2 \ll 1$ and we can safely neglect the contribution of  $Q$ in $\delta N$. 

In conclusion, the only contribution of the gauge field in curvature perturbations comes in
$\delta N$ as given by the second term in  Eq. (\ref{App delta N}).  As a result  we have
\begin{eqnarray}
\label{psiA}
\zeta & = &  \delta N \nonumber\\
~&=& -\frac{\phi}{2 M_P^2} \delta\phi + 2IN\frac{\delta \dot{A_x}}{\dot{A}}.
\end{eqnarray}

We are interested in curvature perturbation power spectrum ${\cal P}_\zeta$ defined
via
\ba
\label{curvaturepower}
\langle \hat{\zeta}_{\vec{k}_{1}}\hat{\zeta}_{\vec{k}_{2}}\rangle = (2 \pi)^3 \delta^3(\vec{k}_{1}+\vec{k}_{2}) { P}_{\zeta}(\vec k_1)
\quad , \quad {\cal P}_{\zeta}(\vec k) = \frac{k_1^3}{2 \pi^2} P_\zeta (\vec k) \, . 
\ea
We decompose the power spectrum into the isotropic part, ${\cal P}_{0}$,
coming from the $\delta \phi$ contribution in Eq. (\ref{psiA}) and the anisotropic power spectrum, $\Delta {\cal P}$,  coming from $\delta \dot A$ in Eq. (\ref{psiA})
\ba
\label{power spectrum}
{\cal P}_{\zeta} \equiv {\cal P}_{0} + \Delta {\cal P} \, .
\ea
As usual the isotropic  power spectrum  is given by
\ba
\label{deltaA}
 {\cal P}_{0} = \frac{H^2}{8\pi^2 M_P^2 \epsilon_{H}} \, .
\ea
To calculate the anisotropic power spectrum we note that $\delta \phi$ and $\delta \dot A$ are
mutually uncorrelated so  $\langle \delta \phi \delta \dot A \rangle |_* =0$. As a results
\ba
\label{gfactor}
\Delta {\cal P} &=& \frac{k_1^3}{2 \pi^2} 4 I^2 N^2 \left\langle  \frac{\delta \dot A_x(k_1)}{A_x}
\frac{\delta \dot A_x(k_2)}{A_x} \right\rangle  \nonumber\\
&=& \frac{k_{1}^3}{2\pi^2} \frac{6IH^2}{\epsilon_{H}k_{1}^3}N^2 \sin^2{\theta} \nonumber\\
&=& 24 \, I N^2 {\cal P}_{0} \sin^2{\theta}
\ea
in which the angle $\theta $ is defined via $\cos \theta = \hat n. \hat k$.
Now comparing this with the anisotropy factor $g_*$ defined via
\ba
{\cal P}_{\zeta}(\vec{k}) = {\cal P}_{0} \left( 1 + g_{*} (\hat k . \hat n)^2  \right) \, .
\ea
we obtained
\ba
\label{g-star}
g_* = -24 I N^2
\ea
Very interestingly this is the result obtained in \cite{Watanabe:2010fh, Bartolo:2012sd} using
the standard in-in formalism.   The advantage of using $\delta N$ formalism is that we only needed to use the background attractor solutions with the information about $ \delta \dot A$ at the time of horizon crossing. This should be compared with the tedious analysis employed in 
\cite{Watanabe:2010fh, Bartolo:2012sd, Emami:2013bk} using in-in formalism to calculate $\Delta {\cal P} $. Physically, one expects that the $\delta N$ method to be applicable in this model. The reason is that all the dynamics between the inflaton field and the gauge field are in the form of local interactions and the dynamics of modes are trivial inside the horizon and at the time of horizon crossing. 

As mentioned in \cite{Watanabe:2010fh, Bartolo:2012sd}, in order to satisfy the observational bound $|g_*| <0.3$, and taking $N=60$ to solve the horizon and the flatness problems,  we require $I < 10^{-6}$. As a result $c$ is very close to unity and
$R\ll1$ as advertised before.

As emphasized in \cite{Bartolo:2012sd} if one allows $N$ to be too large then the 
accumulative anisotropies produced from the IR modes can become too large. Therefore, the total number of e-foldings should not be too large in this model.

%%%%%%%%%%%%%%%%%%%%%%%%%%%%%%%%%%%%%%%%%%%%%%%%%
%%%%%%%%%%%%%%%%%%%%%%%%%%%%%%%%%%%%%%%%%%%%%%%%%

\section{Bispectrum and Trispectrum}
\label{bispectrum}

In this section we calculate the bispectrum and the trispectrum in the anisotropic inflation model studied in previous section using our $\delta N$ method and compare the results with the corresponding results in \cite{Bartolo:2012sd} and \cite{Shiraishi2012} obtained from
the standard in-in formalism. As we shell see the results for bispectrum and the trispectrum
are in exact agreements.

As usual, in order to calculate the bispectrum and the trispectrum in $\delta N$ formalism, we have to expand $\delta N$ to higher orders in perturbations. The expansion of $\delta N$ to
linear order is given in previous section in Eq. (\ref{psiA}). Here we generalize it to second order. To this purpose, we perturb the attractor solution Eq. (\ref{cons-eq}) to second order in fields perturbations
\begin{align}
\label{attractor-per}
\frac{\delta I}{I}&=  \frac{2 f_{,\phi}}{f}\delta \phi +  \frac{2 \delta \dot{A_x}}{\dot{A}} + 
\left[ \bigg{(}\frac{f_{,\phi}}{f}\bigg{)}^2 + \frac{ f_{,\phi \phi}}{f}   \right] \delta \phi^2  + \bigg{(}\frac{\delta \dot{A}}{\dot{A}}\bigg{)}^2 +  \frac{4 f_{,\phi}}{f} \frac{\delta \dot{A_x}}{\dot{A}} \delta \phi  \nonumber\\
& -2 \delta N \bigg{(}1+  \frac{2 f_{,\phi}}{f}\delta \phi +  \frac{2 \delta \dot{A_x}}{\dot{A}}   \bigg{)} +2 \delta N^2
\, .
\end{align}
This formula gives a relation between $\delta I, \delta N$ and different powers of $ \delta \phi$ and  $\delta \dot A$. On the other hand, Eq. (\ref{delta N}) from the perturbation of the evolution of $\phi(\alpha)$  gives a relation between $\delta I, \delta N$ and $\delta \phi$ which is valid to all orders in $\delta I$ and $\delta N$.  Now plugging back Eq.(\ref{attractor-per}) into Eq.(\ref{delta N}) and  keeping the leading corrections from $I \ll 1$ we obtain
\ba
\label{delta N total}
\delta N = N_{\phi} \delta \phi + N_{\dot{A}} \delta \dot{A} + \frac{N_{\phi \phi}}{2} \delta \phi^2  +  \frac{N_{\dot{A} \dot{A}}}{2} \delta \dot{A}^2 + N_{\phi \dot{A}} \delta \phi \delta \dot{A_x}
 \ea
 in which to leading order in $I$
 \ba
 N_\phi \simeq  -\frac{\phi}{2 M_P^2}   \quad , \quad
 N_{\phi \phi} \simeq \frac{2 f_{,\phi}^2}{f^2} + \frac{ 2f_{,\phi \phi}}{f} + \frac{\phi^2}{M_P^4}
 +\frac{ 4  \phi}{M_P^2}  \frac{f_{,\phi}}{f}
 \ea
and
\ba
N_{, \dot A} \simeq \frac{2 I N}{\dot A} \quad , \quad N_{, \dot A \dot A} \simeq \frac{2 I N}{\dot A^2}
\quad , \quad
N_{, \phi \dot A} \simeq \frac{4 I N}{\dot A} \frac{f_\phi}{f}
\ea
Having calculated $\delta N$ to second order in  Eq. (\ref{delta N total}), we can calculate the bispectrum  $B_{\zeta}(\vec k_{1}, \vec k_{2}, \vec k_{3})$ defined via
\ba
\label{bi- def}
\langle \zeta (\vec k_{1}) \zeta (\vec k_{2}) \zeta (\vec k_{3}) \rangle &\equiv& \left( 2 \pi \right)^3 \delta^3 \left( \vec k_{1} +  \vec k_{2} +  \vec k_{3}\right) B_{\zeta}(\vec k_{1}, \vec k_{2}, \vec k_{3}) 
\ea
There are three contributions into bispectrum; (a): $N_{, \phi \phi}N_{,\phi} N_{,\phi} \langle \delta \phi ^4 \rangle$, (b): $N_{, \dot{A}  \dot{A}}N_{, \dot{A}} N_{,\dot{A}} \langle
\delta \dot{A}^4 \rangle$ and (c): $N_{, \phi \dot{A}}N_{, \phi} N_{,\dot{A}}\langle
\delta \phi ^2\delta \dot{A}^2 \rangle$. The term (a) is the one expected from scalar field theory and is very small. The term (b) is purely from the gauge field while the term (c) is from the mixing of inflaton and the gauge field. One expects that the contribution of term (c) to be sub-leading as compared to the contribution of the term (b). Indeed, a direct analysis shows that the ratio of (b) to (c) is $N$ so for $N\sim 60$ one can safely neglect the contribution from the term (c).  In conclusion,  the leading contribution  to the  bispectrum comes from $\langle\delta  \dot A^4 \rangle $ and
\ba
\label{leading three0}
\langle \zeta (k_{1}) \zeta (k_{2}) \zeta (k_{3})  \rangle  &\simeq& \frac{1}{2}
N_{, \dot{A}\dot{A}}( k_{1}) N_{, \dot{A}}( k_{2}) N_{, \dot{A}} ( k_{3}) \int \frac{d^3p}{(2\pi)^3} \langle \delta \dot{A}_{x} (\vec k_{1})  \delta \dot{A}_{x} (\vec k_{2})
\delta \dot{A}_{i} (\vec p) \delta \dot{A}_{i} (\vec k_{3} - \vec p) \rangle + 2 \mathrm{perm.} \nonumber\\
&=& 4 I^3 N(k_{1}) N(k_{2}) N(k_{3}) \int \frac{d^3p}{(2\pi)^3} \langle \delta \dot{A}_{x} (\vec k_{1})  \delta \dot{A}_{x} (\vec k_{2})
\delta \dot{A}_{i} (\vec p) \delta \dot{A}_{i} (\vec k_{3} - \vec p)\rangle + 2 \mathrm{perm.}%\nonumber\\
%&=& 288 I N(k_{1}) N(k_{2}) N(k_{3}) \bigg{(} C(k_{1}, k_{2})P_0(k_{1})P_0(k_{2}) + 2 \mathrm{perm.} \bigg{)} \left( 2 \pi \right)^3 \delta^3 \left( \vec k_{1} +  \vec k_{2} +  \vec k_{3}\right)
\ea
in which $N(k_i)$ represents the time when the mode $k_i$ leaves the horizon.
Now in the Coulomb gauge $A_0=0$, the gauge field perturbations $\delta A_i (\vec k)$ are given by \cite{Bartolo:2012sd}
\ba
\frac{\delta \vec{\dot{A}}}{\dot{A}}|_{t_k} = \sum_{\lambda}\vec{\epsilon}_{\lambda} \frac{\sqrt{3}H}{\sqrt{2I\epsilon_{H}k^3}}
\ea

Plugging these in Eq. (\ref{leading three0}) yields
\ba
\label{leading three}
\langle \zeta (k_{1}) \zeta (k_{2}) \zeta (k_{3})  \rangle  &\simeq&
288 I N(k_{1}) N(k_{2}) N(k_{3}) \bigg{(} C(\vec k_{1}, \vec k_{2})P_0(k_{1})P_0(k_{2}) + 2 \mathrm{perm.} \bigg{)} \left( 2 \pi \right)^3 \delta^3 \left( \vec k_{1} +  \vec k_{2} +  \vec k_{3}\right) \, ,
\ea
in which the momentum shape function $C(\vec k_1, \vec k_2) $ is defined via
\ba
C(\vec k_{1}, \vec k_{2})\equiv\bigg{(}1 -   (\widehat k_1.\widehat n)^2  -   (\widehat k_2.\widehat n)^2 + 
(\widehat k_1.\widehat n) \,  (\widehat k_2.\widehat n) \,  (\widehat k_1.\widehat k_2)  \bigg{)}
\ea
To obtain Eq. (\ref{leading three})
we have used  $P_0{(k_{1})} = \frac{H^2}{4k_{1}^3\epsilon_{H}M_{P}^2}$
for the isotropic power spectrum and
\ba
\label{modeA}
\langle \frac{\delta \dot{A}_{i}(\vec k_{1})}{ \dot{A}} \frac{\delta \dot{A}_{j}(\vec k_{2})}{ \dot{A}} \rangle = \frac{3H^2}{2I\epsilon_{H}k^3M^2_{P}}
\bigg{(} \delta_{ij} - \widehat{k}_{1i}\widehat{k}_{1j} \bigg{)}\left( 2 \pi \right)^3 \delta^3 \left( \vec k_{1} +  \vec k_{2}\right)
\ea
Using  Eq. (\ref{leading three}), one can calculate the bispectrum as
\ba
\label{fnl leading}
B_\zeta(\vec k_1, \vec k_2, \vec k_3)=  288 I N(k_1) N(k_2) N(k_3) \left( C(\vec k_{1}, \vec k_{2}) P_0(k_{1})P_0(k_{2})
+ 2 \mathrm{perm.} \right)
%f_{NL}(\vec k_1, \vec k_2, \vec k_3) = 240 I N(k_{1}) N(k_{2}) N(k_{3})C(k_{1}, k_{2}) \, .
\ea
This completes our results for the bispectrum. As expected, the shape of the bispectrum is anisotropic. Very interestingly our formula Eq. (\ref{leading three}) and Eq. (\ref{fnl leading})
agree exactly with the result of \cite{Bartolo:2012sd} obtained using the standard in-in formalism.

To calculate $f_{NL}$ we go to the squeezed limit $k_{1} \ll k_{2} \simeq k_{3}$ in which
\ba
f_{NL} (\vec k_1, \vec k_2, \vec k_3) = \lim_{k_1 \rightarrow 0} \frac{5}{12}
\frac{B_\zeta(\vec k_1, \vec k_2, \vec k_3)}{P_\zeta(k_1) P_(k_2)} \, .
\ea
In this limit we get
\ba
\label{fnl squeezed}
f_{NL} &=& 240 I N(k_{1}) N(k_{2})^2 C(\vec k_{1}, \vec k_{2})   \quad \quad  \quad
 ( k_{1} \ll k_{2} \simeq k_{3} )  \\
&\simeq& 10 N\,  |g_*| \, C(\vec k_1, \vec k_2)
\ea
Taking $N \sim 60$ and $|g_* | \sim 0.1$ and neglecting the orientation-dependence  in $f_{NL}$ this leads  to large non-Gaussianity $f_{NL} \sim 60$.

Now we are in the position to calculate the trispectrum of our model. The trispectrum is defined via
\ba
\label{taunl}
\langle \zeta (k_{1}) \zeta (k_{2}) \zeta (k_{3}) \zeta (k_{4}) \rangle = (2 \pi)^3 \delta^3 \left( \vec k_{1} +  \vec k_{2} +  \vec k_{3} + \vec k_{4}\right) T_\zeta (\vec k_1, \vec k_2, \vec k_3, \vec k_4) \, .
%\tau_{NL}(k_{1}, k_{2}, k_{3}, k_{4})
%\bigg{(}P(k_{3}) P(k_{4})P(k_{1}+k_{3}) + 11 \mathrm{perm.}  \bigg{)}\left( 2 \pi \right)^3 \delta^3 \left( k_{1} +  k_{2} +  k_{3} + k_{4}\right)
\ea
In the collapsed limit $\vec k_1 + \vec k_3 = \vec k_2 + \vec k_4 =0$
we can calculate the parameter $\tau_{NL}$ via
\ba
\label{tau-NL-def}
\tau_{NK}(\vec k_i) = \lim_{\vec k_1 + \vec k_3\rightarrow 0} \frac{1}{4}
\frac{T_\zeta(\vec k_1, \vec k_2, \vec k_3, \vec k_4)}{P_\zeta( | k_1 + k_3| ) P_\zeta (k_1) P_\zeta (k_3)}
\ea
The trispectrum $\langle \zeta (\vec k_{1}) \zeta ( \vec  k_{2}) \zeta (\vec  k_{3}) \zeta (\vec k_{4}) \rangle$ have
6 contributions in the forms of
$$
N_{, \phi \phi}^2N_{,\phi}^2  \langle \delta\phi^6\rangle  \,\, , \,\,
N_{, \phi \dot{A}}^2N_{,\phi}^2  \langle \delta \phi ^4 \delta \dot{A}^2  \rangle  \,\, , \,\,
N_{, \phi \dot{A}}^2N_{,\dot{A}}^2 \langle \delta \phi ^2 \delta \dot{A}^4 \rangle
\,\, , \,\,
N_{, \phi \dot{A}}N_{, \dot{A} \dot{A}}N_{,\dot{A}} N_{, \phi} \langle \delta \phi^2
\delta \dot{A}^4 \rangle  \,\, , \,\,
N_{, \phi \dot{A}}N_{, \phi \phi}N_{,\dot{A}} N_{, \phi} \langle \delta \phi^4 \delta \dot{A}^2
\rangle
$$
and $N_{,\dot{A} \dot{A}}^2N_{,\dot{A}}^2 \langle \delta \dot{A}^6\rangle$.
As in the case of bispectrum, one can easily check that the last term has the dominant contribution in trispectrum. Therefore  to leading order we have
\ba
\label{leading four}
\langle \zeta (\vec k_{1}) \zeta (\vec k_{2}) \zeta (\vec k_{3}) \zeta (\vec k_{4}) \rangle &\simeq& N_{, \dot{A}\dot{A}}(k_{1})N_{, \dot{A}\dot{A}}(k_{2}) N_{, \dot{A}}(k_{3})N_{, \dot{A}}(k_{4})\int \frac{d^3p}{(2\pi)^3}  \int \frac{d^3q}{(2\pi)^3}
\nonumber\\
&&\left\langle \delta \dot{A}_{x} (\vec k_{3}) \delta \dot{A}_{x} (\vec k_{4})
\delta \dot{A}_{i} (\vec q)\delta \dot{A}_{i} (\vec k_{1}-\vec q) \delta \dot{A}_{j} (\vec p)\delta \dot{A}_{j} (\vec k_{2}-\vec p)
\right\rangle + 5 \mathrm{perm.} \nonumber\\
&=& 3456 I N(k_{1}) N(k_{2}) N(k_{3}) N(k_{4}) \bigg{(}  D(\vec k_{3}, \vec k_{4}, \vec k_{1}+\vec k_{3}  )P(k_{3}) P(k_{4})P(|\vec k_{1}+\vec k_{3}|) \nonumber\\
&&~~+  \mathrm{11 perm.}  \bigg{)}\left( 2 \pi \right)^3 \delta^3 \left( \vec k_{1} +  \vec k_{2} +  \vec k_{3} + \vec k_{4}\right) \, ,
\ea
in which $ D(\vec k_{3}, \vec k_{4}, \vec k_{1}+\vec k_{3}  )$ refers to the trispectrum's shape function and is given by
\ba
\label{four shape}
D(\vec k_{3}, \vec k_{4}, \vec k_{1}+\vec k_{3}) &=& 1 -  (\widehat k_4.\widehat n)^2 
-  (\widehat k_3.\widehat n)^2  - (\widehat{ k_1 + k_3}. \widehat n)^2 +
(\widehat k_3. \widehat n)  (\widehat k_4. \hat n)  (\widehat k_3. \widehat k_4)  +  (\widehat k_4. \widehat n)   (\widehat{ k_1 + k_3}. \widehat n)  (\widehat{ k_1 + k_3}. \widehat k_4)
\nonumber\\
&+&  (\widehat k_3. \widehat n)   (\widehat{ k_1 + k_3}. \widehat n)  (\widehat{ k_1 + k_3}. \widehat k_3) -   (\widehat k_3. \widehat n)    (\widehat k_4. \widehat n)    (\widehat{ k_1 + k_3}. \widehat k_3)  (\widehat{ k_1 + k_3}. \widehat k_4)  \, .
\ea
Comparing Eq. (\ref{four shape}) with the definition of trispectrum we obtain
\ba
\label{T-zeta}
T_\zeta(\vec k_1, \vec k_2, \vec k_3, \vec k_4) =  3456 I N(k_{1}) N(k_{2}) N(k_{3}) N(k_{4}) \bigg{(}  D(\vec k_{3}, \vec k_{4}, \vec k_{1}+\vec k_{3}  )P_\zeta(k_{3}) P_\zeta(k_{4})P_\zeta(|\vec k_{1}+\vec k_{3}|) +  \mathrm{11 perm.}  \bigg{)} \, .
\ea
Now going to the collapsed limit $\vec k_1 + \vec k_3 = \vec k_2 + \vec k_4 =0$
and using the definition of $\tau_{NL}$ given in Eq. (\ref{tau-NL-def}) we obtain
\begin{align}
\label{tauNL collapsed}
\tau_{NL}(k_{1}, k_{2}, k_{3}, k_{4}) \simeq 3456 I N(k_{3})^2 N(k_{4})^2  D(\vec k_{3}, \vec k_{4}, \vec k_{1}+\vec k_{3}  ) \, .
\end{align}
As in the case of bispectrum the trispectrum is anisotropic so $\tau_{NL}$ has 
direction-dependence. Comparing our trispcetrum with the results of \cite{Shiraishi2012} obtained from in-in formalism,  we find the exact agreement between these two results.

Now we are in the position to check the SY inequality between $f_{NL}$ and $\tau_{NL}$ which states 
\ba
\label{SY}
\tau_{NL} \ge \left( \frac{6}{5} f_{NL} \right)^2 \, .
\ea
The importance of SY inequality as a tool to rule out inflationary scenarios was studied
in \cite{Komatsu:2010hc, Sugiyama:2011jt}.  The SY inequality presented in the form of 
Eq. (\ref{SY}) is for the models in which $f_{NL}$ and $\tau_{NL}$ are either scale-invariant or have the same
scale-dependence. In our case, see also \cite{Rodriguez:2013cj, BeltranAlmeida:2011db},  we have complicated shape-dependent for $f_{NL}$ and $\tau_{NL}$ so the original SY inequality as given in Eq. (\ref{SY}) is not applicable.
Instead, in \cite{Assassi:2012zq} a general integral representation of SY is proved which states
\ba
\int d^3 q_1 d^3 q_2 \tau_{NL} (\vec q_1, \vec k - \vec q_1, \vec q_2, -\vec q_2 - \vec k)
P_\zeta(q_1) P_\zeta(q_2) \ge
\left( \int d^3 q \frac{6}{5} f_{NL} (\vec q, -\vec q -\vec k, \vec k) P_\zeta(q)
\right)^2 \, ,
\ea
in which $k\rightarrow 0$. As demonstrated in \cite{Shiraishi2012} this integral from of
SY inequality does hold in our model if we assume $g_* <1$. Indeed, the condition $g_*<1$ 
is necessary from the observational constraints on the power spectrum and the consistency of our starting assumption that the anisotropic power spectrum is sub-leading so  $\Delta { \cal P} < {\cal P}_0$.

To see qualitatively that the SY inequality in its simple form, Eq. (\ref{SY}), does hold in our model, let us neglect the direction-dependence in $f_{NL}$ and $\tau_{NL}$ coming from $D(\vec k_{3}, \vec k_{4}, \vec k_{1}+\vec k_{3}  )$ and $C(\vec k_1, \vec k_2) $. As a result
\begin{align}
\label{ration}
\frac{\tau_{NL} (k_{1}, k_{2}, k_{3}, k_{4})}{(\frac{6}{5})^2f_{NL}(k_{3})f_{NL}(k_{4})} \simeq   \frac{1}{g_*} \,     \quad \quad \quad \quad (g_* < 1)
\end{align}
in which we have taken $N(k_i) = N$  and $g_* \simeq -24 I N^2$ from Eq. (\ref{g-star}).
Demanding that $g_* < 1$ from the cosmological observations and also from the consistency of our analysis we conclude that   $\tau_{NL} > \frac{6}{5} ( f_{NL})^2$ so the SY inequality does hold.

%As a further probe of anisotropic inflation and primordial non-Gaussianity, it would be very interesting to study the effect of anisotropic seed perturbations for large scale structure formation \cite{Shiraishi:2013sv}.

%%%%%%%%%%%%%%%%%%%%%%%%%%%%%%%%%%%%%%%%%%%%%%%%%%
\section{Conclusion and Discussions}
\label{discussions}

In this work we have presented the consistent $\delta N$ formalism in anisotropic backgrounds.
We have demonstrated that the separate universe approach works.  In each homogenized patch 
the local continuity equation and the local Friedmann equation hold which have the same form as the corresponding background equations. We note that the Hubble expansion rate appearing in continuity equation, $H( \mathbf{x}, t)$, and the Hubble expansion rate appearing in Friedmann equation, ${\cal H}( \mathbf{x}, t)$, are different.  

The anisotropic pressure has two different effects in $\delta N$ analysis. One is the direct effect encoded by the term containing $Q$  in Eq. (\ref{Q-eq}). The second effect is indirect and comes from the back-reactions of the source of anisotropic pressure on the dynamics of other fields, such as the inflaton field. We have calculated these two effects in model of anisotropic inflation containing a $U(1)$ gauge field. We have shown that the second effect, i.e. the back-reaction effect, is much larger than the direct effect coming from the $Q$
term in $\delta N$ formula. In previous works in the literature, this back-reaction effect during inflation was not taken into account. The gauge  field contribution $\delta A_i$ is added trivially as a non-interacting field during inflation.

Taking into account the back-reaction of gauge field on inflaton dynamics, we have calculated $\delta N$ to linear and to second order in perturbations. We have demonstrated that our $\delta N$ formalism exactly reproduces the power spectrum and the bispectrum results obtained in previous works using standard in-in formalism. This is a non-trivial verification of the validity of our $\delta N$ analysis. The advantage in using $\delta N$ formalism is that all we need to know to calculate the power spectrum and higher order correlations is the background dynamics and the profile of gauge field fluctuations on super-horizon scales. This method seems to be considerably simpler than the standard in-in formalism.

We also calculated the bispectrum and the trispectrum in anisotropic inflation model. The bispectrum and the trispectrum are both orientation-dependent and scale-dependent. As a result the SY inequality in its simple form is not applicable. However,  a generalization of the SY inequality in its integrated form indeed holds.  

Large amount of non-Gaussianity with 
$f_{NL} \sim 60$ can easily be generated in this model. It is an interesting exercise to compare the predictions of this model on CMB and large scale structure formation along 
\cite{Pullen:2007tu, Shiraishi:2013sv}.

%%%%%%%%%%%%%%%%%%%%%%%%%%%%%%%%%%%%%%%%%%%%%%%%%%
\acknowledgments

We would like to thank Shant Baghram,  Paolo Creminelli,  Eiichiro Komatsu,  Karim Malik, Mohammad Hossein Namjoo, Marco Peloso and Misao Sasaki  for useful discussions and comments. R. E. would like to thank ICTP for hospitality where this work was in its final stage.
We also thank the anonymous referee for the insightful comments about the ordering of
$\delta N$ which were helpful to improve our presentations.

%%%%%%%%%%%%%%%%%%%%%%%%%%%%%%%%%%%%%%%%%%%%%%%%%%
\appendix

\section{Metric perturbations and Gauge invariant perturbations}
\label{gauge-transformations}

In this Appendix we specify the properties of metric transformations under the general coordinate transformation and construct the gauge invariant curvature perturbations $\zeta$ to linear order in perturbation theory $\delta$. 

The background metric is given in Eq. (\ref{Bianchi-metric1}).  The most general form of the scalar perturbations for the Bianchi I metric is introduced in Eqs. \eqref{ADM} and (\ref{gamma-ij}). For  the later convenience we introduce new variables $\tilde{\beta^i}$ and $\tilde{\gamma}^{ij}$ as
\ba
\partial_i \tilde{\beta^i} \equiv \beta^i ,\qquad \partial_i\partial_j\tilde{\gamma}^{ij} \equiv \gamma^{ij},
\ea
with no sum on repeated indices.  

Consider the general coordinate transformation
\ba
\label{xi}
x^\mu \rightarrow x^{\mu} + \xi^{\mu} \quad \quad , \quad \quad
\xi^\mu = \left( \xi^0 \, ,\, \partial_i \hat \xi^i \right)
\ea
in which $\xi^0$ and $ \xi^i = \partial_i \hat \xi^i$ for $i=1,2,3$ are scalars.
Under the coordinate transformation Eq. (\ref{xi}) we have
\ba
\delta g_{\mu \nu} \rightarrow \delta g_{\mu \nu} -\bar{g}_{ \mu \nu, \kappa}\, \xi^{\kappa} -
\bar{g}_{\alpha \nu}\, \partial_\mu \xi^\alpha -\bar{g}_{\alpha \mu}\, \partial_\nu \xi^\alpha
\ea
in which $\bar{g}_{\alpha \mu}$ is the background Bianchi metric given in Eq. (\ref{Bianchi-metric1}).
More explicitly, one can check that
\ba
A && \rightarrow A - \partial_t \xi^0 \\
\tilde{\beta}^i&& \rightarrow \tilde{\beta}^i -\frac{1}{a_i^2} \xi^0 -  \partial_t \hat \xi^i \\
\psi_i && \rightarrow \psi_i - H_i \xi^0-2 \partial_i^2 \hat \xi^i \\
\tilde{\gamma}^{ij} && \rightarrow \tilde{\gamma}^{ij} - \dfrac{a_i}{a_j} \hat \xi^i - \dfrac{a_j}{a_i} \hat \xi^j\\
\ea
in which  ${\cal N} \equiv 1 +A$. 

If we apply the gradient expansion approximation $\partial_i^2= O(\epsilon^2)$, then $\zeta$
defined via
\ba
\label{zeta-def}
-\zeta=\frac{(\psi_1+\psi_2+\psi_3)}{3}- \frac{(H_1+H_2+H_3)}{3}\frac{\delta \rho}{\dot{\rho}} = \psi -H \frac{\delta \rho}{\dot{\rho}}
\ea
is gauge invariant and can be interpreted as the average curvature perturbations in our setup.
The definition of $\zeta$
to all orders of perturbation theory can be found in \cite{Lyth:2004gb}.

%%%%%%%%%%%%%%%%%%%%%%%%%%%%%%%%%%%%%%%%%%%%%%%%%%
\section{Gradient Expansion Ordering of Perturbations}
\label{App-B}

In this section we estimate the ordering of $\beta^i$ and  $\delta q^\mu$ and calculate the contributions of
$\delta q^\mu$ and $\delta \pi^{\mu}_{\nu}$ in the energy conservation equation,
Eq. (\ref{cont}).

First of all let us check the transverse conditions on the heat flow and anisotropic pressure. By definition one has
\ba
\label{trans-cond}
u^{\mu} q_{\mu} =0 \qquad  , \quad  u^{\mu} \pi^{\nu}_{\, \, \mu} =0.
\ea
%Noting that $\beta = {\cal O} (\delta)$ in which $\delta$ as usual is the order of perturbations,
%\footnote{We can consider as \cite{Lyth:2004gb} that $\beta = {\cal O} (\epsilon)$ but that assumption here is not necessary and the above assumption is enough for our purpose.}
The fluid's 4-velocity  can be read as 
\ba
\label{u-up-dn}
u^{\mu}=\left[\dfrac{1}{\cal N}, \vec{0} \right]+{\cal O}(\epsilon^2)  \quad ,\qquad u_{\mu}=\left[-{\cal N}, \dfrac{\beta_i}{\cal N}\right]+{\cal O}(\epsilon^2).
\ea
From the background equations we conclude  that $\bar q^\mu$ and $\bar \pi^{0}_{\mu}$ are zero. Now using the transverse condition \eqref{trans-cond} one concludes that to all order
\ba
\delta q_0 =0 \quad , \quad \delta \pi^{\nu}{}_{0} =0
\ea
This also yields  $\delta q^{0} = {\cal O} (\delta^2)$.
%\ba \label{q0-pi0-order} \delta q^{0} = {\cal O} (\delta^2),\qquad \delta \pi^{\nu}{}_{0}= {\cal O} (\delta^2) \ea
For the ordering of $\delta \pi^{0}{}_{i}$ one has
\ba
\label{pi-0-i}
\delta \pi^{0}{}_{i}= a_i^2 \delta \pi^i{}_{0} + \beta_i \pi^{i}{}_{i}  = \beta_i \pi^{i}{}_{i}  \, .
\ea
We will use this equation later in order to find the ordering of the gradient expansion of perturbations.
 
Let us now look at the gradient expansion ordering of $\beta^{i}$. 
For this we look at $\delta G^{0}{}_{i}$ and $\delta G^i{}_0$ components of Einstein equations. With some efforts one can show that 
\ba
%\label{G-0-i}
\delta G^{0}{}_{i} &=& \epsilon {\cal O} (\delta) + \beta {\cal O}(\delta) 
\\
\nonumber
%\label{G-i-o}
\delta G^i{}_0 &=& \beta^i \left(3 \bar H_i \bar H- \sum_i \bar H_i^2 - 3\dot{\bar H}+ \dot{\bar H}_i  \right) + \epsilon {\cal O} (\delta) + \beta {\cal O}(\delta) 
\\
&=&\beta^i \left(\bar{R}^i{}_{i} -\bar{R}^0{}_{0} \right) + \epsilon {\cal O} (\delta) + \beta {\cal O}(\delta)  .
\ea
The easiest way to see this is to adopt the local inertial frame in which $\Gamma^\alpha_{\beta \gamma} =0$. %In this coordinate system, one can check that $$
Therefore, the corresponding Einstein equations yield
\ba
\label{G-0-i}
  \delta T^{0}{}_{i}&=& \epsilon {\cal O} (\delta) + \beta {\cal O}(\delta) 
\\
\label{G-i-0}
 \delta T ^i{}_{0}&=& \beta^i \left(\bar{R}^i{}_{i} -\bar{R}^0{}_{0} \right) + \epsilon {\cal O} (\delta) + \beta {\cal O}(\delta)  .
\ea
Similarly, using the spatial components of energy momentum conservation one can put limits on $\delta T^{0}{}_i$. The continuity equation $\nabla_{\mu} T^{\mu}_i$ to leading order yields
\ba
\label{T-i-0-con}
\nonumber
(\partial_0 +3H-H_i) \delta T^{0}{}_i -a_i^2 H_i \delta T^{i}{}_0 &=& -H_i \beta_i \left(\bar{T}^i{}_i - \bar{T}^0{}_0  \right) +\epsilon {\cal O} (\delta) + \beta {\cal O}(\delta)
\\
&=&- H_i \beta_i \left(\bar{R}^i{}_i - \bar{R}^0{}_0  \right)+\epsilon {\cal O} (\delta) + \beta {\cal O}(\delta) \, .
\ea
One can show that the expression, $\bar{T}^i{}_i - \bar{T}^0{}_0  = \bar{R}^i{}_i - \bar{R}^0{}_0  $  is non-vanishing in general and is of the order of $\dot{H}$. 

Plugging Eqs. (\ref{G-0-i}) and (\ref{G-i-0}) into continuity equation Eq. (\ref{T-i-0-con}) yields 
 \ba
\label{T-0-i-con}
(\partial_0+3H-H_i) \delta T^{0}{}_{i} &=\epsilon {\cal O} (\delta) + \beta {\cal O}(\delta) .
\ea
This equation shows that  $\delta T^{0}{}_{i}$ has decaying  solutions approximately like $1/a^{2}$. So one can readily deduce that  $\delta T^{0}{}_{i}$ should be higher order in gradient expansion as
\ba
\label{T-0-i-order}
\delta T^{0}{}_{i} &=\epsilon^2 {\cal O} (\delta) + \epsilon \beta {\cal O}(\delta)  \, .
\ea
Before discussing about the consequences of the above equation it is more convenient to rephrase Eq. (\ref{G-i-0}) as follows
\ba
\label{G-i-0-2}
-\delta q_i &=& \beta_i \left(\bar{T}^i{}_{i} -\bar{T}^0{}_{0} \right)   + \epsilon {\cal O} (\delta) + \beta {\cal O}(\delta) ,
\ea
Furthermore  for   $\delta T^{0}{}_{i}$  we have
\ba
\label{G-0-i-2}
\delta T^{0}{}_{i}&=& (\bar{\rho}+\bar{p}) \beta_i -\delta q_i +\delta \pi^{0}{}_{i}+\epsilon {\cal O} (\delta) + \beta {\cal O}(\delta)  
\ea
By using Eq. \eqref{G-i-0-2} to eliminate $\delta q^i$ in favor of $\beta^i$ and Eq. \eqref{pi-0-i} to express $\delta \pi^{0}{}_{i}$ as a function of $\beta^i$  one can show that the leading order terms of $\beta^i$ cancel each other and one obtains
\ba
\label{G-0-i-3}
\delta T^{0}{}_{i}&=& \beta^i \, {\cal O}(\delta)+\epsilon \,{\cal O} (\delta) + \beta\, {\cal O}(\delta)  \, .  
\ea
On the other hand, comparing Eq. (\ref{T-0-i-order}) with Eq. (\ref{G-0-i-3}), one obtains the following result for the ordering of $\beta^i$
\ba
\beta^{i} &={\cal O} (\epsilon) ,
\ea
This also yields 
\ba
\label{ordering}
\delta q_i= {\cal O} (\epsilon) 
\\
\label{ordering2}
\delta \pi^{0}{}_{i}= {\cal O} (\epsilon) 
\ea
Now we investigate the contribution of heat flow in continuity equation Eq. \eqref{cont}. Using Eq. \eqref{ordering} and Eq. \eqref{ordering2} one finds that $\delta q^{\mu} \sim {\cal O} (\epsilon )$ and by noting that the background value of $q^{\mu}$ is also zero, we get
\be
\label{heat-con} 
-u_\mu u^\nu \nabla_\nu q^\mu + \nabla_\mu q^\mu = {\cal O} (\epsilon\delta,\epsilon^2 )
\ee
As a result one can deduce that heat conduction can be ignored in the continuity equation at the first order of perturbations and gradient expansion.
\\
Now it is time to calculate the contribution of anisotropic pressure on the continuity equation
\ba
u^{\mu} \nabla_{\nu} \pi^{\nu}{}_{\mu} &=& \bar{u}^{\mu} \bar{\nabla}_{\nu} \bar{\pi}^{\nu}{}_{\mu} +\delta (u^{\mu} \partial_{\nu} \pi^{\nu}{}_{\mu}+u^{\mu} \Gamma^{\nu}_{\nu \rho} \pi^{\rho}{}_{\mu} - u^{\mu} \Gamma^{\rho}_{\mu \nu } \pi^{\nu}{}_{\rho}) 
\ea
Noting that $u^{\mu}=\left[1/{\cal N}, \vec{0} \right]+{\cal O}(\epsilon^2)$ and  $\delta \pi^{\nu}{}_{0} =0$ to the all orders of perturbations $\delta$, one has
\ba
\label{aniso-pres-con}
u^{\mu} \nabla_{\nu} \pi^{\nu}{}_{\mu} = \dfrac{-1}{\cal N} H_i (\mathbf{x},t)  \pi^i_i (\mathbf{x},t).
\ea

%%%%%%%%%%%%%%%%%%%%%%%%%%%%%%%%%%%%%%%%%%%%%%%%%%

\section{$(i\neq j)$ Components of Einstein equation}
\label{off-Ein}

In this Appendix we look into off-diagonal components of spatial Einstein equations
$M_P^2 \delta G^i_j = \delta T^i_J$ for $i \neq j$. These equations are trivial at the background level. The leading order perturbation equations lead
\ba
\label{gamma-1}
{}^{(1)}\ddot{\gamma}_{ij} + 3H {}^{(1)}\dot{\gamma}_{ij}+ \left[3 H_i H_j-3({\cal H}^2+\dot{H}) -2\sum_{i} H_i^2 - |\epsilon^{ijk}|\dot{H}_k  \right]{}^{(1)} \gamma_{ij} = \dfrac{2}{a_i a_j M_P^2} {}^{(1)}\delta \pi_{ij},\qquad (i\neq j)
\ea
in which $\epsilon^{ijk}$ is the Levi-Civita symbol. As one can see the above equation has decaying solutions. This is due to the fact that the background metric does not admit off-diagonal spatial components.  Weinberg has argued that the anisotropic stress  for a wide class of theories to be some linear combinations of $\delta u$, $\delta p$  and $\delta \rho$ \cite{Weinberg:2003sw}. We partially extend this assumption and  assume that the anisotropic stress can also obtain contributions from gauge fields, $A_{\mu}$. So anisotropic stress tensor $\pi_{ij}$ for $i \neq j$ can be some linear combination of $\partial_i \partial_j p$, $\partial_i \partial_j \rho$, $\partial_i  u_j$, $\partial_i  A_j$, $\partial_i q_j$, $u_i u_j$, $u_i A_j$, $u_i q_j$ and finally $A_i q_j$ . As $\dot{A}_i \dot{A}_j $ for $i \neq j$ is forbidden by the background equations, this term does not contribute to the off-diagonal part of $\pi_{ij}$. These contributions are at least at the first order of gradient expansion $\epsilon$. So one readily deduces that
\ba
\pi_{ij} ={\cal O} (\epsilon).
\ea
Now,  Eq. \eqref{gamma-1} can be rephrased as \ba
\label{gamma-1b}
{}^{(1)}\ddot{\gamma}_{ij} + 3H {}^{(1)}\dot{\gamma}_{ij}+ m^2_{ij} \gamma_{ij} = {\cal O}(\epsilon) \, ,
\ea
 with $m^2_{ij} \sim H^2$ so $\gamma_{ij}$  has decaying solutions scaling  approximately as $a^{3/2}$. This is a consequence of the fact  that the background equations do not admit $\gamma_{ij} \neq 0$ for $i \neq j$. At the second order in perturbation variables $\delta$, the homogeneous equation has the same form, but it can be verified that all  possible source terms  are at least at the first order in gradient expansion $\epsilon$. This argument can be repeated for all orders of perturbations. This argument leads to the conclusion that in the $n$-th order
of perturbation theory, the off diagonal spatial part of metric,  after the decaying solutions become negligible, are of the first order of gradient expansion. As a result, one deduces
\ba
\gamma_{ij} = \cal {O} (\epsilon)
\ea
This equation is important for gradient expansion of Einstein equations.  

%%%%%%%%%%%%%%%%%%%%%%%%%%%%%%%%%%%%%%%%%%%%%%%%%%

\end{document}